\documentclass[preprint]{ptephy_v1}

\usepackage{boites}
\preprintnumber{RESCEU-28/15} 
 
\begin{document}
    
\title{All-sky coherent search for continuous gravitational waves in 6-7 Hz band with a torsion-bar antenna}

\author[1,2]{Kazunari Eda\thanks{eda@resceu.s.u-tokyo.ac.jp}}
\author[3]{Ayaka Shoda}
\author[1]{Yuya Kuwahara}
\author[2]{Yousuke Itoh} 
\author[1,2,3]{Masaki Ando}
\affil[1]{
Department of Physics, 
Graduate School of Science, 
University of Tokyo, Tokyo 113-0033, Japan
}
\affil[2]{
Research center for the early universe (RESCEU), 
Graduate School of Science,   
University of Tokyo, Tokyo 113-0033, Japan
} 
\affil[3]{
Gravitational Wave Project Office, Optical and Infrared Astronomy Division, 
National Astronomical Observatory, Osawa 2-21-1, Mitaka, 
Tokyo 181-8588, Japan
}

\begin{abstract}%
A torsion-bar antenna (TOBA) is a low-frequency terrestrial gravitational wave (GW) antenna 
which consists of two orthogonal bar-shaped test masses. We upgraded the prototype TOBA 
and achieved the strain sensitivity $10^{-10}$ $\text{Hz}^{-1/2}$ at around 1 Hz. 
We operated the upgraded TOBA (called the ``Phase-II TOBA'') located at Tokyo in Japan for 22.5 hours 
and performed an all-sky coherent search for continuous GWs using the $\mathcal{F}$-statistic. 
We place upper limits on continuous GWs from electromagnetically unknown sources in the frequency range from 6 Hz to 7 Hz 
with the first derivative of frequency less than $7.62 \times 10^{-11}$ $\text{Hz}/\text{s}$ 
using data from the TOBA. 
As a result, no significant GW signals are found in the frequency band $6-7$ Hz. 
The most stringent upper limit upper limit on the dimensionless GW strain with 95\% confidence level 
in this band is  $3.6 \times 10^{-12}$ at 6.84 Hz. 
\end{abstract}

\subjectindex{F34}

\maketitle


\section{Introduction}\label{Sec:intro}
The first direct detection of gravitational waves (GWs) are anticipated 
in half a decade by large-scale laser interferometric GW detectors 
such as the advanced Laser Interferometer
Gravitational wave Observatory (advanced LIGO) \cite{Abbott:2007kv}, 
Advanced Virgo \cite{Accadia:2012zzb}, and KAGRA \cite{Aso:2013eba}. 
A network of these advanced ground-based GW detectors will 
reach unprecedented sensitivity which may be enough to establish GW astronomy. 
One of the most promising targets for them is continuous GW from 
a rapidly spinning neutron star (NS) 
which is generated due to non-axisymmetry around its spin axis.  
Detection of continuous GWs from pulsars would shed light on NS equations 
of state via the observed GW amplitude. 

There are two kinds of efforts to search for GWs from rapidly rotating
isolated neutron stars. One is to search for GWs from electromagnetically known
pulsars and the other, sometimes called a blind search, tries to find GWs
in a wide parameter space where the parameters include the source sky direction, the GW emission frequency,
and its time derivative. Detection of GWs from an electromagnetically known
pulsar would enable us to reveal the GW generation mechanism 
via the relation between the NS's spin and the GW frequency. Detections of
GWs from many pulsars in blind searches would statistically tell us the
beaming angle of pulsars.  

To date, data from the initial LIGO and Virgo science runs have been used to 
place upper limits on GW amplitudes from unknown isolated pulsars with GW frequencies above 20 Hz 
\cite{Abbott:2005pu, Abbott:2006vg, Abbott:2007td, Abbott:2008uq, Abbott:2008rg, 
Abbott:2009nc, Abadie:2011wj, Aasi:2012fw, Aasi:2013lva, Aasi:2014mtf, Aasi:2015rar}.  
On the other hand, continuous GWs below 20 Hz have yet to be investigated 
because seismic noise on the Earth hinders the sensitivities of detectors to GWs in
such a low-frequency regime.
One of the solutions to avoid the noise due to the ground motion is to construct GW detectors formed by 
satellites in space such as 
the evolved Laser Interferometer Space Antenna (eLISA) \cite{AmaroSeoane:2012je} and 
the DECi-hertz Interferometer Gravitational wave Observatory (DECIGO) \cite{Kawamura:2011zz}. 
Another solution is to devise detector configurations on the Earth 
such as the torsion-bar antenna (TOBA) \cite{Ando:2010zz}, 
atomic interferometers \cite{Dimopoulos:2008sv}, 
the juggled interferometer \cite{Friedrich:2014dua}, 
and the full-tensor detector \cite{Harms:2015zaa}. 
Indeed, the Australia Telescope National Facility catalogue lists about 1500 pulsars in the frequency
range from 1 Hz to 10 Hz, while it contains only about 400 above 10 Hz
\cite{Manchester:2004bp}. 
Hence, it is interesting to explore the low-frequency regime, 
although the expected GW amplitude scales as frequency squared
 
Here, we first search for unknown continuous GWs coherently 
in the low-frequency regime using data from an upgraded TOBA (called the ``Phase-II TOBA'').  
A TOBA is a ground-based low-frequency GW antenna which is 
composed of two orthogonal bar-shaped test masses. 
When a GW passes through the antenna the two bars rotate differentially around their centers. 
GW signals can be extracted by monitoring the bar rotations using laser interferometers. 
So far, the prototype TOBA was used to set upper limits
on the abundance of the stochastic GW background \cite{Ishidoshiro:2011ii, Shoda:2013oya}. 
In this paper, we report on the results of an all-sky search for 
continuous GWs using the upgraded TOBA data. 

This paper is organized as follows. 
Section \ref{Sec:observation} the presents detection mechanism of 
a TOBA and summarizes the experiments with the upgraded TOBA. 
The adopted method of data analysis and its results are described in Sec. \ref{Sec:data_analysis}. 
Conclusion and summary are given in Sec. \ref{Sec:conclusion}.

\section{Observation}\label{Sec:observation}
\subsection{Detection mechanism}
A TOBA is a ground-based antenna for low-frequency GWs, 
and was originally proposed in Ref. \cite{Ando:2010zz}. 
A TOBA consists of two bar shaped orthogonal test masses 
to which mirrors are attached at both ends . 
The two bars rotate differentially due to the passage of incident GWs. 
The GW signals can be read in the following way. 
An input laser beam is split into two orthogonal beams at a beam splitter.  
The two beams are reflected by the mirrors attached to the ends of the bars 
and are recombined at the beam splitter. 
The GW signals can be obtained by measuring the optical path differences 
at a photo-detector placed in a different direction from the laser. 
The angular motions of the bars are written as follows:  
\begin{align}
  I\ddot{\theta} \left(t\right) + \gamma \dot{\theta} \left(t\right) + \kappa \theta \left(t\right) = \dfrac{1}{4} \ddot{h}_{jk} q^{jk}
\end{align}
where we denote the moment of inertia, the dissipation coefficient, the spring constant, and 
the dynamical quadrupole moment of the two bars by $I$, $\gamma$, $\kappa$, and $q$, respectively.  
A TOBA is sensitive to incoming GWs above the resonant frequency 
$f_{\text{res}} = \sqrt{\kappa/I}/2\pi$, which can be set to be below 1 Hz.

\subsection{Phase-II TOBA}
In previous works, we have constructed a prototype TOBA 
which is composed of a single 20 cm test mass and has succeeded 
in putting constraints on the abundance of stochastic GWs \cite{Ishidoshiro:2011ii, Shoda:2013oya}.
We have developed the Phase-II TOBA based on Ref. \cite{Eda:2014rha}. 
The main features of the Phase-II TOBA are common-mode noise rejection, 
the multi-output system, and the active and passive vibration systems. 
The Phase-II TOBA has two 24 cm bar-shaped test masses each of which is suspended 
by two parallel tungsten wires near its center. 
In order to reduce the common-mode noise effectively, the two test masses are installed 
in such a way that their centers of mass are positioned at the same point on the horizontal plane.
The motions of the bars in both the horizontal and vertical planes are monitored 
by using fiber Michelson laser interferometers, 
so that we can obtain three independent output signals unlike the
previous prototype TOBA. 
We introduced a hexapod-type active vibration isolation system 
to reduce the seismic noise at around 1 Hz; see Ref. \cite{Shoda:2015} for more details. 
    
\begin{figure}[htbp] 
\centering  
\includegraphics[width=8cm,clip]{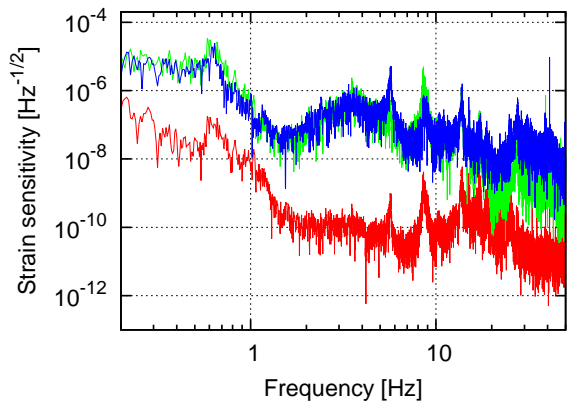}
\caption{\label{Fig:TOBANoiseCurve}
The strain sensitivity curve of the Phase-II TOBA.
The horizontal axis shows the frequency and 
the vertical axis shows the square root of the noise spectral density $\sqrt{S_n\left(f\right)}$.  
The red, blue, and green lines correspond to 
the output signals from the $xy$, $xz$, and $yz$ degrees of freedom, respectively. 
}
\end{figure}

\begin{figure}[htbp] 
\centering  
\includegraphics[width=7cm,clip]{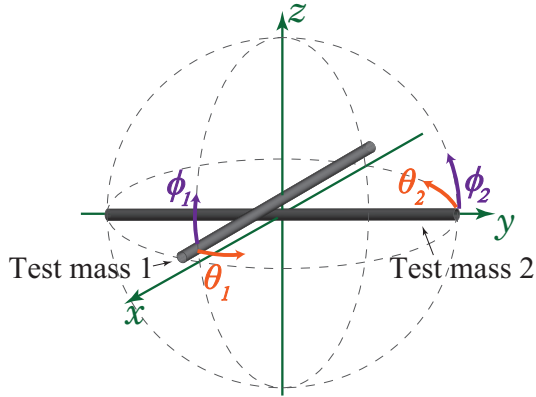}
\caption{\label{Fig:TOBA}
Layout of the Phase-II TOBA. The two bar-shaped test masses can move independently 
in the directions of the $x$, $y$, and $z$ axes. 
The three independent output signals can be obtained by monitoring 
the bar motion on the $xy$, $xz$, and $yz$ planes. 
}
\end{figure}

We placed the Phase-II TOBA in Tokyo
($35^{\circ} 42' 49.0'' \text{N}$, $139^{\circ} 45' 47.0'' \text{E}$)
and operated it for 22.5 hours from 6:18 UTC, December 11, 2014 to 4:48 UTC, December 12, 2014. 
The measured strain sensitivity of the Phase-II TOBA is shown in Fig. \ref{Fig:TOBANoiseCurve}
in which the red, blue, and, green lines correspond to 
the output signals from the $xy$, $xz$, and $yz$ degrees of freedom, respectively. 
The $z$ axis is in the local vertical direction, while the $x$ and $y$ axes align with the two bars
when they are at rest (see also Fig. \ref{Fig:TOBA}).
We achieved the GW equivalent-strain sensitivity $10^{-10}$ $\text{Hz}^{-1/2}$ 
at around 1 Hz for the signal on the horizontal plane. 
The sensitivity is limited by the seismic noise below 2 Hz and by unexpected noise in the optical fiber above 2 Hz. 
The peaks appearing at around 0.7 Hz, 5.7 Hz, 8.5 Hz, and 14 Hz
correspond to the resonance of the optical bench, and
the resonance of the vibration isolation table in the directions of $y$, $x$, and $z$ axes, respectively. 
It should be noted that we do not incorporate the data obtained from monitoring the vertical planes into our analysis 
because their sensitivities are unfortunately much worse than the
sensitivity from the horizontal plane, as is evident from Fig. \ref{Fig:TOBANoiseCurve}. 
 
\subsection{Statistical properties of the data}\label{subsec:data_properties}
Searches for continuous GWs are effectively carried out 
using short time baseline Fourier transforms (SFTs) of the data.
The length of each SFT segment is determined by taking into account 
the effects of stationarity of the data, 
the spinning of the Earth, and the spin-down of the GW source. 
We set the SFT length to be 9,000 seconds for these reasons. 
It should be noted that the SFT baseline length can  be taken longer at 
lower frequencies as long as the data stationarity is assured. 

\begin{figure}[htbp] 
\centering 
\includegraphics[width=17cm,clip]{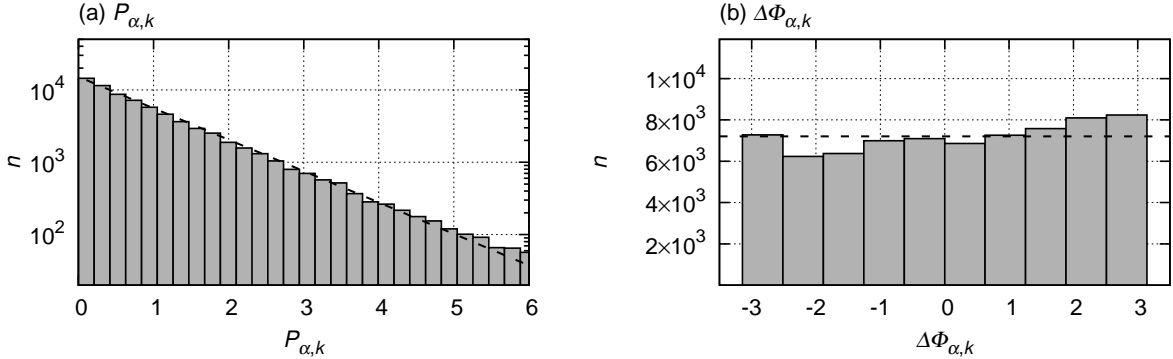}
\caption{\label{Fig:StatProp}
Histogram of (a) the power $P_{\alpha,k}$ and (b) the phase
 $\Delta \Phi_{\alpha, k}$ of the TOBA SFT data at the $k$th frequency bin. 
If the data is distributed according to a Gaussian distribution, 
$P_{\alpha, k}$ lies on the straight line described by a dotted line. 
If the data is stationary, 
$\Delta \Phi_{\alpha,k}$ is distributed uniformly in the range of $\left[-\pi, \pi \right]$
as described by a dotted line.
}
\end{figure}

We investigated the statistical properties of our data in the band 
6-7 Hz using SFTs. It is convenient to define the following quantity 
to evaluate to what extent the noise obeys a Gaussian distribution \cite{Abbott:2003yq}. 
\begin{align}
  P_{\alpha, k} = \dfrac{\left|\tilde{x}_{\alpha, k}\right|^2}{\langle \left|\tilde{x}_{\alpha, k}\right|^2 \rangle}
\end{align} 
where $\tilde{x}_{\alpha, k}$ denotes the SFT data at the frequency bin $k$ of the $\alpha$th SFT segment and 
$\langle \cdot \rangle$ denotes the ensemble average over the $\alpha$th SFT. 
$P_{\alpha, k}$ can be regarded as the normalized noise power in the
frequency bin $k$. 
The histogram of $P_{\alpha, k}$ is shown in the left panel of Fig. \ref{Fig:StatProp}. 
If the data is distributed according to a Gaussian distribution, 
$P_{\alpha, k}$ is proportional to an exponential, 
or in other words, $P_{\alpha, k}$ is aligned with 
a straight line in a semi-log plot. 
The measured values of the mean and the standard deviation are 1.00 and 1.08, respectively. 
Thus, we can regard that our data follows almost a Gaussian distribution. 
We also studied the stationarity of the data by computing the difference between 
adjacent phases of the SFT data \cite{Abbott:2003yq}, 
\begin{align}
 \Delta \Phi_{\alpha, k} = \Phi_{\alpha, k} - \Phi_{\alpha, k-1}. 
\end{align} 
If the data is stationary, $\Delta \Phi_{\alpha,k}$ obeys a unifom distribution in the range of $\left[-\pi , \pi \right]$. 
The histogram of measured $\Delta \Phi_{\alpha,k}$ is shown in the right panel of Fig. \ref{Fig:StatProp} 
in which strong non-stationarity is not found. 

\section{Data analysis}\label{Sec:data_analysis}
\subsection{GW signal}
A pulsar is a rapidly rotating NS whose spin frequency is nearly constant, say $f_0$. 
GWs from such a source are generated by a non-axisymmetry of the pulsar around its spin axis. 
The GW amplitude is expressed by 
\begin{align}
 &h_0 = \dfrac{16 \pi^2 G}{c^4 r} \varepsilon I f_0^2
 \label{Eq:h0}
\end{align} 
where $G$, $c$, and $I$ are Newton's gravitational constant, 
the speed of light, and the NS's moment of inertia, respectively. 
The non-axisymmetry is characterized by the parameter $\varepsilon$ called ellipticity. 
The GW singal we expect to detect from a rapidly rotating NS is described as 
\begin{subequations}
\begin{align}
  &h \left(t\right) = h_+ \left(t\right) F_+ \left(t\right) +  h_{\times} \left(t\right) F_{\times} \left(t\right), \label{Eq:h} \\
  &h_+\left(t\right) = h_0 \dfrac{1+\cos^2\iota}{2} \cos \Phi \left(t\right), \label{Eq:hplus} \\
  &h_{\times} \left(t\right) = h_0 \cos \iota \sin \Phi \left(t\right) \label{Eq:hcross}
\end{align}
\end{subequations}
where $h_0$ is the overall amplitude,  
$\Phi\left(t\right)$ is the GW phase measured at the solar system barycenter (SSB), 
and $\iota$ is the inclination which is the angle between the line of sight and the spin axis.
The antenna pattern functions $F_+ \left(t\right) $ and $F_{\times} \left(t\right)$  
represent the response of the antenna 
to the plus and cross polarization modes of the incoming GWs. 
The antenna patterns of the TOBA rotated by $45^{\circ}$ on the antenna plane 
are identical to that of a $90^{\circ}$ interferometer \cite{Eda:2014rha}. 
The spin of the Earth around its axis gives rise to the amplitude modulation 
which is described by the time dependence in $F_+$ and $F_{\times}$. 
  
The Earth's spin around its axis and the Earth's rotation around the Sun  
bring Doppler modulation to the GW phase up to the first derivative of frequency 
as follows:  
\begin{subequations}
\begin{align}
  &\Phi\left(t\right) = \phi_0 + 2 \pi \Delta t \hat{f} \left(\Delta t\right) , \label{Eq:Phi} \\
  &\hat{f} \left(\Delta t\right) = f_0 + \dfrac{1}{2} \dot{f} \Delta t,  \label{Eq:hatf} \\
  &\Delta t = \tau + \dfrac{\boldsymbol{r}_d \cdot\boldsymbol{n}}{c} + \Delta_{\text{rel}}   - t_0 \label{Eq:deltaT}
\end{align}
\end{subequations}
where $\phi_0$, $\tau$, $\boldsymbol{r}_d$, and $\boldsymbol{n}$ 
denote the initial phase at the reference time $t_0$, the arrival time of the GW measured at the detector, 
the detector position on the Earth with respect to the SSB,
and the unit vector pointing toward the NS from the SSB. 
The unit vector $\boldsymbol{n}$ is related to the equatorial coordinates:
right ascension $\alpha$ and declination $\delta$. 
The timing correction $\Delta_{\text{rel}}$ represents relativistic effects such as 
the Einstein delay and the Shapiro delay.

\subsection{$\mathcal{F}$-statistic}
We use a detection statistic called the $\mathcal{F}$-statistic 
to discriminate whether or not an expected GW signal exists in the data \cite{Jaranowski:1998qm}. 
The $\mathcal{F}$-statistic is derived from the method of maximum likelihood function 
and is known as the most powerful test from a frequentist standpoint 
according to the Neyman-Pearson lemma \cite{Prix:2009tq}.  
On the stationary Gaussian noise assumption, the log-likelihood function is expressed as 
\begin{align}
 \ln \Lambda = \left(x | h \right) - \dfrac{1}{2} \left(h|h\right)
\end{align}
where $\left( \cdot | \cdot \right)$ denotes the noise-weighted inner
product defined as 
\begin{align}
 \left(x | y \right) = 4\text{Re} \int_{0}^{\infty} \dfrac{\tilde{x}\left(f\right) \tilde{y}^{\ast} \left(f \right)}{S_n \left(f\right)} df. 
\end{align}
The maximization of $\ln \Lambda$ over the amplitude parameters 
$\boldsymbol{\lambda} = \left\{h_0, \cos \iota, \psi, \phi_0 \right\}$ 
leads to the $\mathcal{F}$-statistic, 
\begin{align}
 2 \mathcal{F} = \underset{\boldsymbol{\lambda}}{\text{max}} \left[ 2 \ln \Lambda \right]. \label{Eq:Fstatistic}
\end{align} 
The number of search parameters is reduced from eight to four in this process. 
In the presence of a GW signal, $2\mathcal{F}$ obeys a 
non-central $\chi^2$ distribution  with four degrees of freedom and 
a non-centrality parameter $\rho^2$, where $\rho$ is the 
average signal-to-noise ratio (SNR) in the case of a signal 
perfectly matched with the template.  
In the absence of any GW signal, $2\mathcal{F}$ obeys the $\chi^2$ distribution function with four degrees of freedom. 
The SNR of the signal is related to the expected value of the $\mathcal{F}$-statistic 
by $\langle 2\mathcal{F} \rangle = 4 + \rho^2$. We set the threshold of $2\mathcal{F}$ to be 68 which 
corresponds to $\rho = 8$.  
  
If the measured value of the $\mathcal{F}$-statistic is below the predetermined threshold, 
we move on to the step of placing a constraint on the GW amplitude $h_0$. 
An upper limit of the amplitude can be defined as a function of the confidence level $C$, $h_0 \left(C\right)$. 
The inverse of $h_0 \left(C\right) $ is written as 
\begin{align}
 C \left(h_0 \right) 
 = \int_{2\mathcal{F}_{\text{obs}}}^{\infty} p \left(2\mathcal{F} | h_0  \right) d\left(2\mathcal{F}\right) 
 \label{Eq:confidence_of_h0}
\end{align}
where $\mathcal{F}_{\text{obs}}$  denotes the observed value of $\mathcal{F}$-statistic and 
$p \left(2\mathcal{F} | h_0 \right)$ denotes the probability distribution function 
of 2$\mathcal{F}$ in the presence of a signal with amplitude $h_0$. 
The value of the upper limit is evaluated by solving Eq.(\ref{Eq:confidence_of_h0}) via 
Monte-Carlo simulations over the unknown parameters $\left\{ h_0, \cos \iota, \psi \right\}$. 
Note that the $\mathcal{F}$-statistic is independent of $\phi_0$.

\subsection{Analysis and results}
Equations (\ref{Eq:h0})-(\ref{Eq:deltaT}) indicate that a continuous GW is characterized by eight parameters when 
we take into account up to the first derivative of frequency. 
The four amplitude parameters are projected out by using the $\mathcal{F}$-statistic.  
Then, the parameters to be searched over become only the phase parameters $\{\alpha, \delta, f_0, \dot{f} \}$. 
The spacing of frequency bins for the templates is chosen by the inverse of twice the observation time.
The grid spacings on the sky positions are chosen such that the maximum mismatch is less than 0.02.
We take both $\Delta \alpha$ and $\Delta \delta$ to be 0.01 radians conservatively.
To reduce the computational burden, we did not search over spindown parameters.  
So our analysis is valid only for NSs with a spindown $\dot{f}$ less
than $1/(2T_{\rm obs}^2) \simeq 7.62\times 10^{-11}$ Hz/s where $T_{\rm obs}
= 22.5$ hours is the data length. 

We make SFTs of 22.5 hours contiguous data by employing lalapps\_MakeSFTs   
in the LIGO scientific collaboration analysis library (LAL) code \cite{LAL:2015}. 
Each segment is windowed by a Tukey window prior to computing the SFTs. 
The length of each SFT segment is chosen as 9,000 seconds 
for the reasons described in Sec. \ref{subsec:data_properties}.
The frequency range to be searched is a 1 Hz band in 6-7 Hz 
where our antenna has good sensitivity. 
The statistical properties of the data in this band are 
described in Sec. \ref{subsec:data_properties}.

\begin{figure}[htbp]
\centering 
\includegraphics[width=17cm,clip]{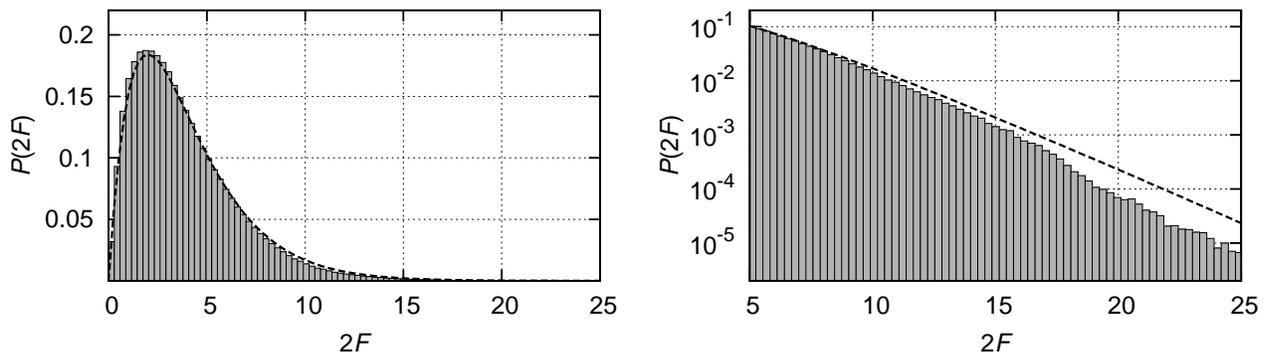}
\caption{\label{Fig:PDF_twoF}
Probability distributions of the $\mathcal{F}$-statistic over 6.10 - 6.11 Hz. 
The right panel is identical to the left panel apart from the scale of the 
vertical axis and the range of the horizontal axis.  
The gray filled areas in both panels represent histograms obtained from the observations. 
In each panel, a dotted line represents a central $\chi^2$ distribution with four degrees of freedom. 
When the data is dominated by Gaussian noise, 
the histogram obeys the dotted line. 
}     
\end{figure}

We compute $2\mathcal{F}$ by making use of lalapps\_ComputeFstatistic\_v2 in the LAL code.
The left panel in Fig. \ref{Fig:PDF_twoF} shows the distribution of 
$2\mathcal{F}$ over a 0.01 Hz band between 6.10 Hz and 6.11 Hz.
The experimentally measured distribution of the $\mathcal{F}$-statistic is represented by 
the gray filled boxes.  The theoretically expected distribution in the case of 
the Gaussian noise is represented by a $\chi^2$ distribution with four degrees of freedom, 
which are given by the dotted line. 
As can be seen in the left panel of Fig. \ref{Fig:PDF_twoF}, the two are agreed with each other very well. 
This indicates that the data we observed is filled with almost Gaussian noise. 
The right panel is identical to the left panel but changes 
the scale of the vertical axis to a the semi-log scale.  
Because of the small non-Gaussian noise, the gray filled area is deviated slightly from the dotted line 
for larger values of the $\mathcal{F}$-statistic. 

\begin{figure}[htbp] 
\centering 
\includegraphics[width=17cm,clip]{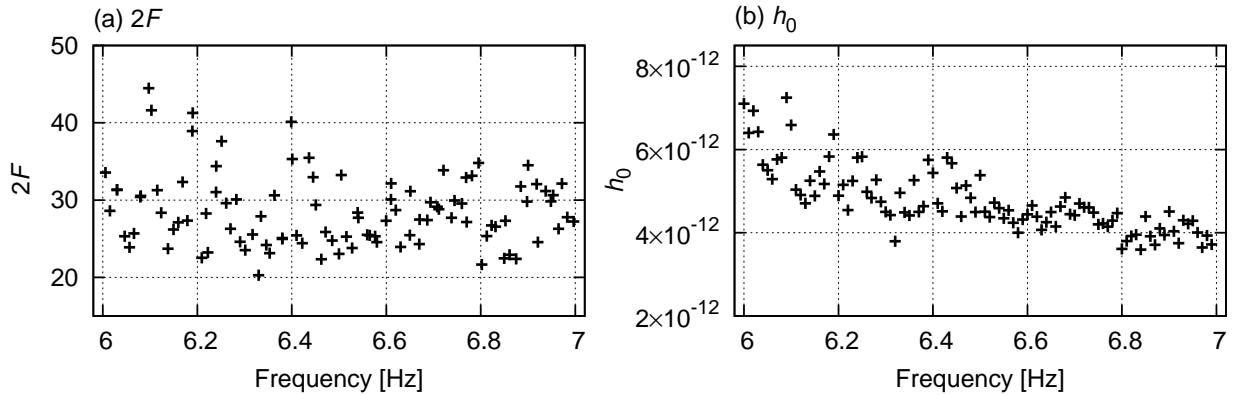}
\caption{\label{Fig:twoF_UL}
(a) Loudest values of $2\mathcal{F}$ in sub-bands of 0.01 Hz width.  
(b) Upper limits on $h_0$ with 95\% confidence level in sub-bands as a function of frequency. 
}
\end{figure}

We divide the 1 Hz band between 6 Hz and 7 Hz into 100 sub-bands each of length 0.01 Hz. 
The loudest values of $2\mathcal{F}$ in each sub-band 
resulting from the all-sky search are computed and are shown in the left panel of Fig. \ref{Fig:twoF_UL}. 
There is no significant candidate whose value of the $\mathcal{F}$-statistic is above the predetermined threshold 
$2\mathcal{F} = 68$ corresponding to $\text{SNR} = 8$. 
Then, we move on to the step of finding the upper limits on $h_0$ by employing Eq. (\ref{Eq:confidence_of_h0}). 
The right panel in Fig. \ref{Fig:twoF_UL} represents the upper limit of $h_0$ with a 95\% confidence level in each sub-band.
The constraints on $h_0$ become tighter as the frequency increases.
This feature basically reflects the noise curve shown in Fig. \ref{Fig:TOBANoiseCurve}. 
The most stringent upper limits on $h_0$ is $3.6 \times 10^{-12}$ at $6.84$ Hz.  
   

\subsection{Discussion}
We can interpret our upper limits on the strain amplitudes in terms of
upper limits on the ellipticity $\varepsilon $ using Eq. (\ref{Eq:h0}).
For instance, when we consider a NS with its moment of inertia $I = 10^{38}$ kg m$^2$ 
at a distance of $r=0.1$ kpc, the most stringent upper limit we obtained 
corresponds to the constraint on the ellipticity of $\varepsilon = 1.7 \times 10^{9}$. 
The maximum possible value of ellipticity is typically of the order of less than $10^{-6}$ \cite{Horowitz:2009ya}, 
so this limit has yet to reach an interesting parameter region. 

One of the proposed configurations of the TOBA [17]
may achieve the best sensitivity of $\sim 10^{-20}$ Hz$^{-1/2}$
at around 0.1 Hz. 
With the proposed TOBA, we can detect GWs from
inspiralling compact binaries such as NS/NS binaries within the Local Group or 
intermediate-mass black hole binaries within 10 Gpc, 
in addition to low-frequency continuous GWs from rapidly rotating
compact stars. In fact, the ${\cal F}$-statistic search method
we employed in this paper 
can be used for inspiralling compact binaries long before their
coalescences \cite{Krolak:2004xp,Prix:2007zh,Whelan:2008zz, Whelan:2009jr}.

\section{Conclusion}\label{Sec:conclusion}
In this paper, we carried out an all-sky search for 
continuous GWs from isolated spinning NSs in 
the frequency range from 6 Hz to 7 Hz using the $\mathcal{F}$-statistic. 
The data was obtained from a 22.5-hour observation with the Phase-II TOBA at Tokyo in Japan 
and has good sensitivity of the order of 1 Hz. 
We converted our data into 9,000-second SFT segments 
and searched coherently for an isolated NSs for all sky positions by using the $\mathcal{F}$-statistic.
As a result, no significant candidates were found at 6-7 Hz 
and the most stringent upper limits on $h_0$ with 95\% confidence level in this band 
is $3.6 \times 10^{-12}$ at 6.84 Hz.

\ack

We thank Nobuyuki Kanda and Koh Ueno for useful comments. 
K. E. and A. S are supported by Grant-in-Aid from 
Japan Society for the Promotion of Science (JSPS), 
JSPS Fellows Grant No. 26.8636 (K. E.) and 24.7531 (A. S.).
This work is supported by JSPS Grants-in-Aid for Scientific Research (KAKENHI)
Grant No. 24244031 (M. A.), 25800126, 15K05070, and 
the MEXT KAKENHI Grant Number 24103005 (Y. I.).
We employed the LIGO Scientific Collaboration Algorithm Library (LAL) in our analysis. 
Computations in this paper were mainly conducted on the ORION 
computer cluster of Osaka City University.

\bibliography{TOBA_CW}

\begin{thebibliography}{10}

\bibitem{Abbott:2007kv}
B.P. Abbott et~al., Rept.Prog.Phys., {\bf 72}, 076901 (2009).

\bibitem{Accadia:2012zzb}
T.~Accadia et~al., JINST, {\bf 7}, P03012 (2012).

\bibitem{Aso:2013eba}
Yoichi Aso et~al., Phys.Rev., {\bf D88}(4), 043007 (2013).

\bibitem{Abbott:2005pu}
B.~Abbott et~al., Phys. Rev., {\bf D72}, 102004 (2005).

\bibitem{Abbott:2006vg}
B.~Abbott et~al., Phys. Rev., {\bf D76}, 082001 (2007).

\bibitem{Abbott:2007td}
B.~Abbott et~al., Phys. Rev., {\bf D77}, 022001, [Erratum: Phys.
  Rev.D80,129904(2009)] (2008).

\bibitem{Abbott:2008uq}
B.~Abbott et~al., Phys. Rev., {\bf D79}, 022001 (2009).

\bibitem{Abbott:2008rg}
B.~Abbott et~al., Phys. Rev. Lett., {\bf 102}, 111102 (2009).

\bibitem{Abbott:2009nc}
B.~P. Abbott et~al., Phys. Rev., {\bf D80}, 042003 (2009).

\bibitem{Abadie:2011wj}
J.~Abadie et~al., Phys. Rev., {\bf D85}, 022001 (2012).

\bibitem{Aasi:2012fw}
J.~Aasi et~al., Phys. Rev., {\bf D87}(4), 042001 (2013).

\bibitem{Aasi:2013lva}
J.~Aasi et~al., Class. Quant. Grav., {\bf 31}, 085014 (2014).

\bibitem{Aasi:2014mtf}
J.~Aasi et~al., Class. Quant. Grav., {\bf 31}, 165014 (2014).

\bibitem{Aasi:2015rar}
J.~Aasi et~al., 1510.03621 (2015).

\bibitem{AmaroSeoane:2012je}
Pau Amaro-Seoane, Sofiane Aoudia, Stanislav Babak, Pierre Binetruy, Emanuele
  Berti, et~al., Class.Quant.Grav., {\bf 29}, 124016 (2012).

\bibitem{Kawamura:2011zz}
Seiji Kawamura et~al., Class. Quant. Grav., {\bf 28}, 094011 (2011).

\bibitem{Ando:2010zz}
Masaki Ando, Koji Ishidoshiro, Kazuhiro Yamamoto, Kent Yagi, Wataru Kokuyama,
  et~al., Phys.Rev.Lett., {\bf 105}, 161101 (2010).

\bibitem{Dimopoulos:2008sv}
Savas Dimopoulos, Peter~W. Graham, Jason~M. Hogan, Mark~A. Kasevich, and
  Surjeet Rajendran, Phys. Rev., {\bf D78}, 122002 (2008).

\bibitem{Friedrich:2014dua}
D.~Friedrich, M.~Nakano, H.~Kawamura, Y.~Yamanaka, S.~Hirobayashi, and
  S.~Kawamura, Class. Quant. Grav., {\bf 31}(24), 245006 (2014).

\bibitem{Harms:2015zaa}
Jan Harms and Ho~Jung Paik, Phys. Rev., {\bf D92}(2), 022001 (2015).

\bibitem{Manchester:2004bp}
R~N Manchester, G~B Hobbs, A~Teoh, and M~Hobbs, Astron. J., {\bf 129}, 1993
  (2005).

\bibitem{Ishidoshiro:2011ii}
Koji Ishidoshiro, Masaki Ando, Akiteru Takamori, Hirotaka Takahashi, Kenshi
  Okada, et~al., Phys.Rev.Lett., {\bf 106}, 161101 (2011).

\bibitem{Shoda:2013oya}
Ayaka Shoda, Masaki Ando, Koji Ishidoshiro, Kenshi Okada, Wataru Kokuyama,
  et~al., Phys.Rev., {\bf D89}, 027101 (2014).

\bibitem{Eda:2014rha}
Kazunari Eda, Ayaka Shoda, Yousuke Itoh, and Masaki Ando, Phys. Rev., {\bf
  D90}(6), 064039 (2014).

\bibitem{Shoda:2015}
A.~Shoda,
\newblock {\em Development of a high-resolution antenna for low-frequency
  gravitational wave observation},
\newblock PhD thesis, University of Tokyo (2015).

\bibitem{Abbott:2003yq}
B.~Abbott et~al., Phys. Rev., {\bf D69}, 082004 (2004).

\bibitem{Jaranowski:1998qm}
Piotr Jaranowski, Andrzej Krolak, and Bernard~F. Schutz, Phys.Rev., {\bf D58},
  063001 (1998).

\bibitem{Prix:2009tq}
Reinhard Prix and Badri Krishnan, Class. Quant. Grav., {\bf 26}, 204013 (2009).

\bibitem{LAL:2015}
{LIGO Scientific Collaboration},
\newblock {LAL{\slash}LALApps: FreeSoftware (GPL) tools for data-analysis},
\newblock http://www.lsc-group.phys.uwm.edu/daswg/ (LALSuite-6.30 released on 4
  August 2015).

\bibitem{Horowitz:2009ya}
C.~J. Horowitz and Kai Kadau, Phys. Rev. Lett., {\bf 102}, 191102 (2009).

\bibitem{Krolak:2004xp}
Andrzej Krolak, Massimo Tinto, and Michele Vallisneri, Phys. Rev., {\bf D70},
  022003, [Erratum: Phys. Rev.D76,069901(2007)] (2004).

\bibitem{Prix:2007zh}
Reinhard Prix and John~T. Whelan, Class. Quant. Grav., {\bf 24}, S565--S574
  (2007).

\bibitem{Whelan:2008zz}
John~T. Whelan, Reinhard Prix, and Deepak Khurana, Class. Quant. Grav., {\bf
  25}, 184029 (2008).

\bibitem{Whelan:2009jr}
John~T. Whelan, Reinhard Prix, and Deepak Khurana, Class. Quant. Grav., {\bf
  27}, 055010 (2010).

\end{thebibliography}

\end{document}